\documentclass[trackchanges]{aastex631} 

\usepackage{acronym, xspace}
\usepackage{booktabs} 
\usepackage{xurl}
\usepackage{amsmath} 

\newcommand{\pastro}{\ensuremath{p_\mathrm{astro}}\xspace{}}

\newcommand{\SoF}{Simons Society of Fellows, Simons Foundation, New York, NY 10010, USA}
\newcommand{\Columbia}{Department of Astronomy and Columbia Astrophysics Laboratory, Columbia University, 550 W 120th St, New York, NY 10027, USA}
\newcommand{\JHU}{William H. Miller III Department of Physics and Astronomy, Johns Hopkins University, Baltimore, Maryland 21218, USA}
\newcommand{\AstroAI}{AstroAI at the Center for Astrophysics \textbar{} Harvard \& Smithsonian, 60 Garden St., Cambridge, MA 02138, USA}
\newcommand{\NU}{Center for Interdisciplinary Exploration and Research in Astrophysics, Northwestern University, 1800 Sherman Avenue, Evanston, Illinois 60201, USA}
\newcommand{\CIT}{Department of Physics, California Institute of Technology, Pasadena, California 91125, USA}
\newcommand{\CITLab}{LIGO Laboratory, California Institute of Technology, Pasadena, California 91125, USA}

\acrodef{NSNS}{binary neutron star}
\acrodef{BHBH}{binary black hole}
\acrodef{BHNS}{black hole--neutron star}
\acrodef{GW}{gravitational wave}
\acrodef{DCO}{double compact object}
\acrodef{LVK}{LIGO-Virgo-KAGRA}
\acrodef{SNR}{signal-to-noise ratio}
\acrodef{FAR}{false-alarm rate}

\begin{document}

\title{Visualizing the Number of Existing and Future Gravitational--Wave Detections \\ from Merging Double Compact Objects}

\author[0000-0002-4421-4962]{Floor S. Broekgaarden}
\affiliation{\SoF}
\affiliation{\Columbia}
\affiliation{\JHU}
\affiliation{\AstroAI}
\email{fsb2127@columbia.edu}
\author[0000-0001-7852-7484]{Sharan Banagiri}
\affiliation{\NU}
\email{sharan.banagiri@u.northwestern.edu}

\author[0000-0003-4507-8373]{Ethan Payne}
\affiliation{\CIT}
\affiliation{\CITLab}
\email{epayne@caltech.edu}



\begin{abstract}
How many gravitational-wave observations from double compact object mergers have we seen to date? This seemingly simple question surprisingly yields a somewhat ambiguous answer that depends on the chosen data-analysis pipeline, {detection threshold} and {other} underlying assumptions. 
To illustrate this we provide visualizations of the number of existing detections from double compact object mergers by the end of the third observing run (O3) based on recent results from the literature. {Additionally, we visualize the expected number of observations from future-generation detectors, highlighting the possibility of up to} millions of detections per year by the time next-generation ground-based  detectors like Cosmic Explorer and Einstein Telescope come online.
We present a publicly available code that highlights the exponential growth in gravitational-wave observations in the coming decades and the exciting prospects of gravitational-wave (astro)physics.
\end{abstract}



\section{Introduction} \label{sec:intro}
Since the first detection of \acp{GW} from a binary black hole merger in 2015 \citep{GW150914:2016detection}, \ac{GW} observatories Advanced LIGO \citep{TheLIGOScientificDetector:2014jea} and Advanced Virgo \citep{Virgo:2015} have rapidly increased the number of observed double-compact object mergers. 
The exact number of confident detections from double-compact object mergers to date, however, depends on the data-analysis pipelines used as well as underlying assumptions such as the detection or signal-to-noise threshold. 
For future observation runs and next-generation detectors, different expected detection numbers can arise from different assumptions on the detector capabilities or the underlying population properties of double compact object mergers in our Universe. 

{As we will discuss, there are a number of feasible choices for claiming the \textit{bona fide} detection of a \ac{GW} signal. We therefore spend the remainder of this manuscript} shedding light on the number of detections in existing and future \ac{GW} observation runs {in two different ways.
First, we outline a pedagogical overview of important concepts regarding the number of detected \ac{GW} events in catalogs before exploring different choices present in the literature and their consequences.
We additionally discuss the future growth of \ac{GW} catalogs as we enter later generations of \ac{GW} detector networks, and present a publicly available code to visualize the number of \ac{GW} sources for outreach and research purposes.
We conclude with an outlook beyond the intricacies of event counting in the coming decades of future detector networks.}

\section{Introduction to number of detections and catalog construction}\label{sec:intro-to-detection-number-and-catalog-construction}

{When determining detection significance, a number of possible measures and criteria exist --- a number of which have been adopted over the last decade.
The most straight-forward calculation is to utilize the matched-filter \ac{SNR}~\citep{Cutler:1994ys}, }
\begin{equation}
    \rho_{\rm mf}^2 = \frac{(4\,{\rm Re} \int_0^\infty df \tilde{\mu}^*(f)\tilde{h}(f)/S_n(f))^2}{4\,{\rm Re} \int_0^\infty df |\tilde{\mu}(f)|^2/S_n(f)}
\end{equation}
{to quantify a detection.
Here, $\tilde{h}(f)$ is the Fourier transform of the measured strain, $\mu(f)$ is the Fourier transform of the template signal model and $S_n (f)$ is the power-spectral density of the interferometric noise.
In the limit of Gaussian noise and ``perfect'' signal templates spanning the entire parameter space, the matched-filter \ac{SNR} is the optimal detection statistic.
Additionally, the noise distribution of \acp{SNR} follows a reduced $\chi^2$ distribution~\citep{Allen:2005fk}  -- providing an analytical expression of the distribution for the significance of an excursion from Gaussianity. 
However, \ac{GW} signals are not the only signal present to cause non-Gaussianity~\citep[e.g.,][]{KAGRA:2020agh, LIGO:2021ppb}.
Glitches due to instrumental noise, seismic motion, and light scattering (to name a few), reduce the viability of a simple \ac{SNR} threshold when analyzing real \ac{GW} data (though such thresholds are often used to estimate detection probabilities when running a full search pipeline is infeasible, e.g.,~\citealt[][]{LIGOScientific:2010nhs}). 
This is the motivating factor behind many of the flagship searches, where significant effort is invested in determining the behavior of the background of noise transients. 
We point to~\cite{LVC:2019a} for a guide to LVK data treatment and analysis methods.}

{Real search algorithms~\citep[e.g.,][]{Klimenko:2016, Messick:2017, Nitz:2018, Sachdev:2019, Aubin:2021}\footnote{See \url{https://emfollow.docs.ligo.org/userguide/analysis/searches.html} for more background information on GW search pipelines.} therefore employ different data-reduction techniques while analyzing the \ac{GW} and output triggers which could potentially be  statistically significant. Associated with each trigger is some ranking statistic $x$. The searches then compute the likelihood ratio between the probability of a trigger being attributed to a \ac{GW} signal and the probability of being a noise transient, which can then be translated into a \ac{FAR}. The \ac{FAR} is a frequentist statistic -- a close analogue of the \textit{p-value} -- which measures the frequency at which something as signal-like as the trigger can be generated by instrumental noise alone. A lower \ac{FAR} hence implies a trigger that is statistically more significant.}

{\ac{FAR} is a statistic that is particularly well-suited for the first detections, when the population of sources has not entirely come into focus. When we have several confident events that have mapped out regions of the parameter space, it is desirable that we also fold in information on the true alarm rate of the signals as well. The statistic developed to fold in this information is called \pastro{}, defined as~\citep{FGMC}}, 
\begin{equation}
    \pastro(x) = \frac{p(x | \mathcal{S} ) \pi(\mathcal{S}) }{p( x| \mathcal{S}) \pi(\mathcal{S}) +p(x|\mathcal{N}) \pi(\mathcal{N})},
\end{equation}
{where $x$ is the trigger and $\mathcal{S}$ and $\mathcal{N}$ stand for the signal and noise hypotheses, respectively. 
Therefore, \pastro{} is simply the probability that the trigger is astrophysical. 
We note that $\mathcal{S}$ assumes not just a signal model (e.g. compact object mergers) but also a specific distribution of sources as that impacts the likelihood of triggers under the signal hypothesis.} 

{Using \pastro{} allows us to distinguish between ``vanilla" events which are expected and those that stand out as outliers when building population statistics. 
It also allows for a more apples-to-apples comparison between events that occur at different observing runs and hence at differing sensitivities (see Fig. 6 and the discussion in Sec. IV C in \citealt{Abbott:2021GWTC3}). 
On the other hand, since \pastro{} depends on some signal model $\mathcal{S}$, in regions of the parameter space with a sparse set of detections, it could simply return back our prior assumptions in the population model. 
As a conservative choice in such cases, one can fall back on using a \ac{FAR} threshold that is low enough to ensure minimal contamination. Once a few detections are obtained in such a region, the signal distribution can be inferred from them.}

{Irrespective of the statistic of choice, constructing a catalog requires defining an arbitrary separation between what is considered a detection and what is not. In cases where high confidence is required, it has been conventional in physics and astronomy to use a $5\sigma$ threshold on the \textit{p value}, corresponding to a probability of 1 in $3 \times 10^{-7}$. With the \ac{GW} community, one common standard is to use a $\pastro = 0.5$ threshold as the dividing line to pick candidates for detailed analysis~\citep{GW150914:rates, Abbott:2021GWTC1,  Abbott:2021GWTC3, Abbott:2021-GWTC-2-1}. }
{Whatever the choice of the threshold, there would be the possibility of a false alarm in the number of events that pass it. Another way to count the number of detections is to simply sum the \pastro{} values, whose mean should converge to the total number of detections, 
as was done, for instance, in Fig. 2 of the GWTC-2.1 paper \citep{Abbott:2021-GWTC-2-1}. }

{In addition to detection, classification of candidates into source types also requires adopting some (population-inclusive) signal model. Catalogs usually calculate probabilities of different sources-types, which can be ambiguous for some candidates. In such cases, we adopt the source type with the maximum probability for the visualizations used in this paper.}

\section{Current Detections from O1, O2, and O3}\label{ch8-sec:O1O2O3}
\begin{figure*}
    \centering
    \includegraphics[width=1\textwidth]{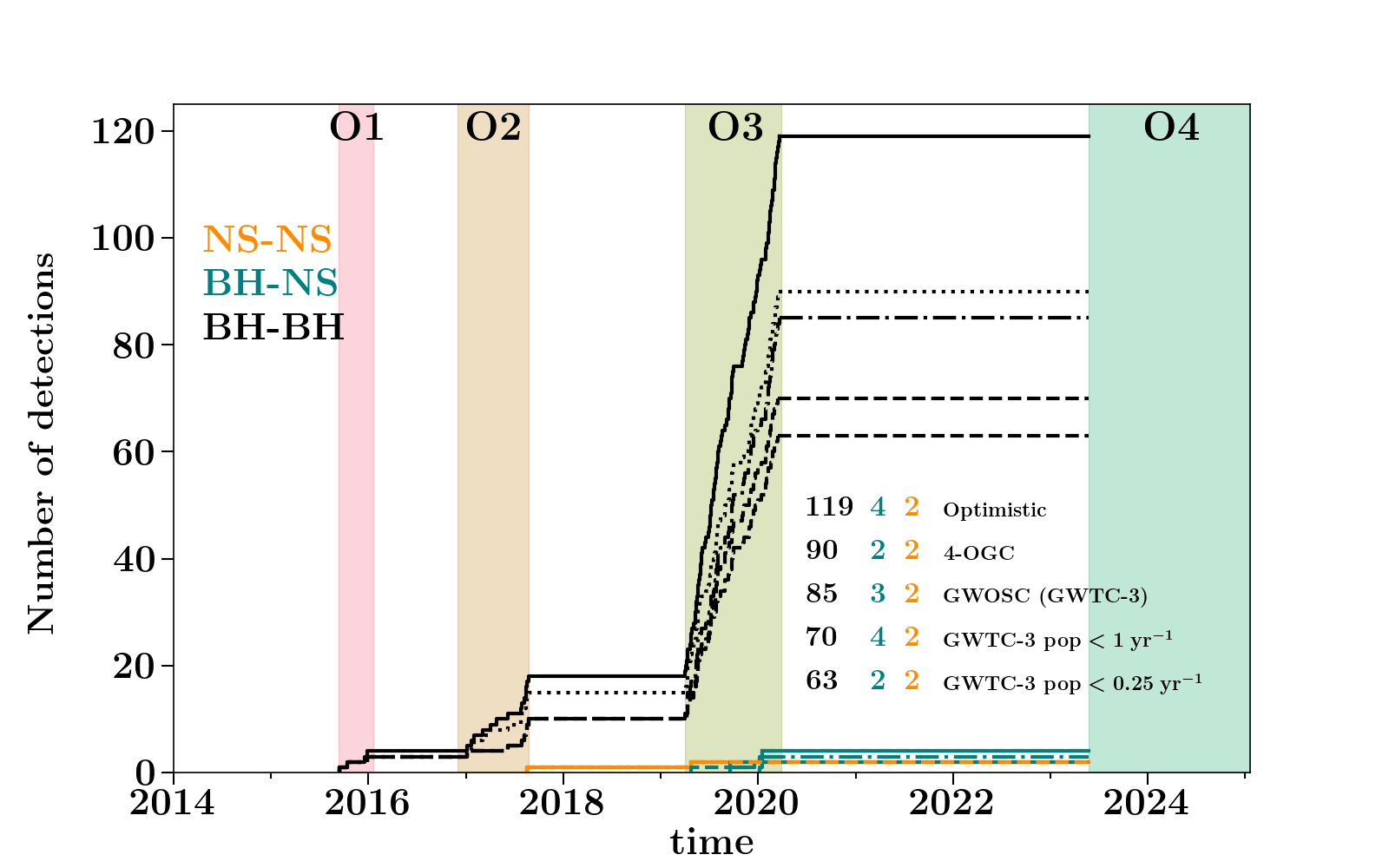} 
    \caption{Cumulative histogram of the number of detected mergers from binary black holes (BH-BH; black), black hole-neutron stars (BHNS; teal), and binary neutron stars (NS-NS; orange) for different data analysis pipelines in the literature from Table~\ref{table:detectionsSummary}. The exact number of detections depends on the underlying assumptions such as the chosen threshold of assigning gravitational-wave events to be of (likely) astrophysical origin.  {The animation shows the number of detections as a function of time and is available at \citet{zenodo:GWvisualizations}.}}
    \label{fig-ch8:known-detections-different-pipelines}
\end{figure*}

At the time of writing, the \ac{LVK} Collaboration has completed and shared the data and detections from observing runs O1, O2, O3a, and O3b in the Gravitational-wave Transient Catalog (GWTC) releases GWTC-1 \citep{Abbott:2021GWTC1}, GWTC-2 \citep{GWTC2}, GWTC-2.1 \citep{Abbott:2021-GWTC-2-1}, and GWTC-3 \citep{Abbott:2021GWTC3}. 
O1 ran from September 12th, 2015 to January 19th, 2016, O2 ran from November 30th, 2016 to August 25th, 2017,  O3a ran from 1 April 2019 until 1 October 2019, and O3b ran between 1 November 2019 and 27 March 2020. Based on the data in these finished runs, several catalogs have been presented in the literature with different numbers of events, based on different underlying assumptions for the detection pipelines and thresholds. We discuss the main catalog options considered in the literature below and summarize them in Table~\ref{table:detectionsSummary} and Figure~\ref{fig-ch8:known-detections-different-pipelines}.

\begin{table*}
\caption{Summary of reported detection numbers after O3b. Acronyms are: FAR (false alarm rate), GWTC (Gravitational Wave Transient Catalog), GWOSC (Gravitational Wave Open Science Center), OGC (Open Gravitational-wave Catalog), IAS (Institute for Advanced Study at Princeton). \\
References: \\
\footnotesize{(a) LVK GWTC events with $p_{\rm{astro}} >0.5$ from the GWOSC \citep{GWOSC:dataset} \edit1{based on GWTC-3}, see \url{https://www.gw-openscience.org/eventapi/html/GWTC/}.}\\ \footnotesize{(b): Table 1 from LVK population paper \citep{Abbott:2021GWTC3pop}, retrieved from the file $\text{gwtc-3\_pop\_events\_table\_1}$  in \citet{Zenodo:GWTC-3pop}.} \\ 
\footnotesize{(c) All detections with $p_{\rm{astro}} >0.5$ from Table~3 in  4-OGC \citep{Nitz:2023-4-OGC}.}\\ 
\footnotesize{(d) Union of GWOSC, OGC, and IAS events. This includes all events found with $p_{\rm{astro}}\geq 0.5$ in the independent searches by \citet{Venumadhav:2019, Venumadhav:2020, 2019PhRvD.100b3007Z, Nitz:2020, Nitz:2021ApJ...922...76N, Nitz:2023-4-OGC, Olsen:2022, mehta2023new, Wadekar:2023}.} \\
\footnotesize{$^{*}$ Note that the GWTC-3 population events technically do not represent a new catalog, but select a \edit1{subset of events from GWTC-3} for the specific analysis of the GW population properties.
} }
\label{table:detectionsSummary}
	\centering
\begin{tabular}{@{}lccccc@{}}
	\toprule
	Label & Total &  BHBH & BHNS & NSNS & Ref \\ 		
	    %
	\hline \hline
	{\it \bf Detections from O1, O2, O3}  &  &  &  &  & \\
	\hline
        %
        {LVK GWTC}                    & \textbf{90}  & \textbf{85} & \textbf{3} & \textbf{2}      & {(a)} \\
        \hline 	
       {GWTC--3 population $\rm{events}^{*}$ with FAR $< 0.25\,\rm{yr}^{-1}$}      & \textbf{67}   & \textbf{63} & \textbf{2}  & \textbf{2}    & {(b)}  \\
         %
        \hline 
       {GWTC--3 population $\rm{events}^{*}$ with FAR  $< 1\,\rm{yr}^{-1}$}         & \textbf{76}   & \textbf{69} & \textbf{4} & \textbf{2}     & {(b)}\\
        \hline 
       {4-OGC with $p_{\rm{astro}} \geq 0.5$}     & \textbf{94}   & \textbf{90} & \textbf{2} & \textbf{2}     & {(c)}\\
        \hline 
       {Optimistic (LVK GWTC, OGC, and IAS events with $p_{\rm{astro}} \geq 0.5$)}     & \textbf{125}   & \textbf{119} & \textbf{4} & \textbf{2}     & {(d)}\\
	%
	\hline 
	%
	%
	%
\bottomrule
\end{tabular}
\end{table*}


%
%

\subsection{LVK {GWTC}} 

The \ac{LVK} Collaboration published several cumulative event catalogs based on observing runs O1, O2, and O3 using a number of search pipelines. {As the field of \ac{GW} astronomy evolved and the number of detections increased, these catalogs were presented using different statistics and thresholds. In GWTC-1, which contains candidates from both O1 and O2 runs, a preliminary FAR threshold of 1 per 30 days in at least one search pipeline and a $\pastro > 0.5$ threshold were applied.
GWTC-1 presented 3 \ac{BHBH} mergers in O1,  one detection of a \ac{NSNS} in O2, and 7 more \ac{BHBH} mergers in O2 which passed this threshold~\citep{Abbott:2021GWTC1}.}

{In GWTC-2, the LVK reported first a total of 39 compact object coalescence detection candidates from the O3a run, including the second \ac{NSNS} merger, using a FAR threshold of 2 $\text{year}^{-1}$ in at least one pipeline~\citep{GWTC2}. While no separate \pastro{} threshold was used, the LVK calculated the astrophysical probability of all \ac{BHBH} candidates and found that they were greater than 50\%. The LVK collaboration reported an updated, deeper list of candidates from O3a in GWTC-2.1 using a FAR threshold of 2 $\text{day}^{-1}$ in at least one pipeline~\citep{Abbott:2021-GWTC-2-1}. While 1201 candidates passed this FAR threshold, only 44 of them had a $\pastro{} > 0.5$, including one \ac{NSNS}. In GWTC-2.1, LVK found that three \ac{GW} events from GWTC-2 fall now below a $\pastro{} > 0.5$ threshold but eight additional candidates were added, due to improved calibration, change of threshold, and subtraction of excess noise.

{In GWTC-3, the LVK Collaboration presented an update of the GWTC with detections from O3b. They found 1083 \ac{GW} candidates which passed a preliminary FAR threshold of 2 $\text{day}^{-1}$ of which 35 have a $\pastro{}>0.5$, including three \acp{BHNS}. 
All in all, this gives a total of 90 events, with 85 \acp{BHBH}, 3 \acp{BHNS}, and 2 \acp{NSNS} that have an astrophysical probability $>0.5$. 
The combined LVK GWTC is presented and kept up-to-date as the GWTC by the Gravitational Wave Open Science Center (GWOSC) \citep{GWOSC:2023paper, GWOSC:dataset}\footnote{As listed on the GWOSC website \url{https://www.gw-openscience.org/eventapi/html/GWTC/} where we used the version from Zenodo by \citet{GWOSC:dataset}.}. Noteworthy, while the \ac{BHNS} merger GW200105\_162426 is included in the publication presenting the first two detections of a \ac{BHNS} by the LVK Collaboration \citep{Abbott:2021-first-NSBH} as well as the GWTC-3 population inference catalogs (which use a \ac{FAR} instead of a \pastro{} threshold; see below), it was found to have an astrophysical probability of $<0.5$. Future \ac{BHNS} detections will help characterize the properties of the \ac{BHNS} population and might address whether GW200105\_162426 is likely of 
astrophysical origin. 
We also note that in the LVK {GWTC} (and all other catalogs mentioned below), GW190814 and GW200210\_092254 with secondary masses of $m_{\rm{2}} \approx 2.5-3$ are most likely classified as \acp{BHBH} as the secondary masses are above the hypothesized maximum mass of a neutron star \citep{  Essick:2020, Huang:2020, Most:2020, Shao:2020, Tsokaros:2020, Godzieba:2021, 
Lim:2021, Tews:2021}. For this reason we also classify both as BHBHs for the visualizations, but the possibility exists that they are instead \ac{BHNS} sources \citep{Abbott:2020gw190814, Abbott:2021GWTC3pop, Abbott:2021GWTC3}.}

{It is common that \ac{GW}-related  studies use subsets of the LVK GWTC, and in the context of the number of \ac{GW} detections two subsets presented in the \ac{LVK} population analysis  \citet{Abbott:2021GWTC3pop} are often used in the literature.}
{First}, in \citet{Abbott:2021GWTC3pop}, the \ac{LVK} Collaboration presents the population properties of compact object mergers based on the data from O1, O2, and O3. The authors use an {sub}set of events {from the LVK GWTC} based on the chosen false-alarm threshold of $< 0.25\,\rm{yr}^{-1}$ to assure  high purity, particularly among mergers with a neutron star\footnote{Where the low-significance event GW190531 was not
included, lacking parameter inference results for use in hierarchical analyses~\citep{Abbott:2021GWTC3pop}:  GW190531  was found in GWTC-2.1 with a FAR of $0.41 \,\rm{yr}^{-1}$ in the GstLAL pipeline, passing the threshold of a FAR of less than 2 per year, but had $\pastro <0.5$, so no parameter inference was performed \citep{Abbott:2021-GWTC-2-1}, leading to its exclusion in the LVK population analysis even though this uses a FAR threshold of $1 \, \rm{yr}^{-1}$ \citep{Abbott:2021GWTC3pop}.}. 
This omits several candidates of moderate significance including {GW190403$\_$051519 and GW200220$\_$061928}. At this threshold LVK expects approximately one event to not be of astrophysical origin.  They present the population catalog in their Table~1, consisting of 63 \ac{BHBH}, 2 \ac{BHNS}, and 2 \ac{NSNS} mergers.
Second, in the same paper the authors instead also discusses the inclusion of all events with FAR $<1\,\rm{yr}^{-1}$, which is a slightly less strict threshold. At this threshold LVK expects approximately 4.6 events to not be of astrophysical origin. Using this threshold, LVK presents this population in Table~1 with 76 events consisting of 70 \ac{BHBH}, 4 \ac{BHNS}, and 2 \ac{NSNS} mergers \citep{Abbott:2021GWTC3pop}. This population catalog is used by the authors to analyze the redshift evolution of \ac{BHBH} mergers.

\subsection{4-OGC: independent catalog events with $p_{\rm{astro}} >0.5$}
In \citet{Nitz:2023-4-OGC}, the authors present an independent data analysis of the O1, O2, and O3 data with slightly different data analysis pipelines.
Specifically, both the LVK analyses and {\citet{Nitz:2023-4-OGC} use the {\tt \sc PyCBC} open-source package, albeit with different settings, such as template bank construction.}
Compared to the earlier OGC catalogs (1-OGC, 2-OGC, and 3-OGC), the authors find seven more \ac{BHBH} events compared to GWTC-3, and also find that three earlier \ac{BHBH} events now pass the threshold for detection, based on updated assumptions. We take all the detections with $p_{\rm{astro}} >0.5$ (which all have an inverse false alarm rate $>100\,\rm{years}$ -- an additional threshold mentioned by the authors) from Table~2  in \citet{Nitz:2023-4-OGC}, which includes a total of 94 observed events with 90 \ac{BHBH}, 2 \ac{BHNS}\footnote{Namely, GW200105\_162426 and GW200115\_042309.}, and 2 \ac{NSNS} mergers. %

\subsection{Optimistic catalog that combines events from the LVK GWTC, OGC and IAS analyses with $p_{\rm{astro}} >0.5$)}
{In addition to the catalogs presented by the LVK Collaboration and by the OGC Collaboration, the IAS Collaboration has also developed independent analysis of the public LIGO-Virgo data and presented additional events. The IAS pipeline focuses on specializing their pipeline to find new events instead of publishing full catalogs. }
For the purpose of visualization, we decide to make an `optimistic' catalog that consists of all events from the LVK GWTC (GWOSC),  4-OGC, and IAS studies, that have an astrophysical origin of $>0.5$. 

This includes in O1 one additional \ac{BHBH} event\footnote{Where we note that this event, GW151205, with $p_{\rm{astro}}=0.53$ was reported by \citep{Nitz:2020} but was found to have $p_{\rm{astro}}<0.5$ in the updated 4-OGC catalog \citet{Nitz:2023-4-OGC} and is omitted by us from the optimistic catalog.} with astrophysical probability $p_{\rm{astro}}\sim 0.71$ \citep{Venumadhav:2019, 2019PhRvD.100b3007Z}. In O2 three additional \ac{BHBH} events with likely astrophysical origin ($p_{\rm{astro}}\geq 0.98$) have been reported by \citet{Venumadhav:2020}. %
These events were found independently with $p_{\rm{astro}} \geq 0.7$ in \citet{Nitz:2020}. 
Three other \ac{BHBH} events with astrophysical probabilities between $0.5$--$0.8$ were also reported in \citet{Venumadhav:2020}, but with astrophysical probabilities $< 0.25$ by 2-OGC in \citep{Nitz:2020}, but then updated to astrophysical probabilities of 0.37, 0.86, and 0.72 in  \citet{Nitz:2023-4-OGC}. 
In \citet{Nitz:2023-4-OGC}, the authors presented with 4-OGC seven additional events with $p_{\rm{astro}} \geq 0.5$ which we also add to the catalog. We also include GW170817A, which has been reported to have an astrophysical probability of 0.86 \citep{Zackay:2021}. 
The same study found the potential gravitational-wave source GWC170402, but we do not include it, as the event cannot fully be described by a binary black hole waveform {(see App. F of~\cite{Abbott:2021GWTC3} for an extended discussion \edit1{on finding events not described by a compact binary coalescence waveform})}.  
We also do not include the intermediate mass binary black hole candidate GW170502, which is considered a marginal trigger in all literature as far as we are aware.\footnote{See the discussion on GW170502 in \citealt{2020ApJ...900...80U}, and Table I and the discussion in \citealt[][]{LVK-IMBH-O2-search:2019}; where the authors reported GW170502 and nine other events as the most significant events in their search for intermediate mass black hole binaries in data from the first and second observing runs.} 

{In \citet{Olsen:2022} and \citet{mehta2023new} the authors re-analyze the Hanford and Livingston coincident data and find 10, and 5 new BHBH events with $p_{\rm{astro}} > 0.5$, in the O3a and O3b data, respectively.   } 
Among these, their events GW190704\_104834, GW190910\_012619, and GW200316\_235947 could also be BHNS events, but we include them here as BHBH sources.
In \citet{Wadekar:2023}, the authors analyze LIGO-Virgo O3 data using an improved IAS pipeline that includes higher harmonics in the \ac{GW} templates and down-weighting noise transients to improve the search sensitivity to high-mass and high-redshift merger events. They find 14 new BHBH events with $p_{\rm{astro}} \geq 0.5$ that we include. 
We refer readers to the tables in, e.g., \citet{Olsen:2022,mehta2023new, Wadekar:2023} for  overviews of the different $p_{\rm{astro}}$ values between the OGC, GWTC and IAS analysis. 

We combine all additional events mentioned in this paragraph with the LVK GWTC (GWOSC) and 4-OGC catalogs to create a new `Optimistic' catalog containing a total of 125 events with 119 \ac{BHBH}, 4 \ac{BHNS}, and 2 \ac{NSNS}.

\section{Future Detections}\label{ch8-sec:future-detections}

We show a visualization of the possible expected observations in the coming decades in Figure~\ref{fig-ch8:expected-detections-different-pipelines}. We emphasize that these are very rough estimates for the detection number and uncertainty. We provide publicly available code that allows the user to change these assumptions. For outreach purposes, it can be useful to show a video with a linear y-axis {because it can be difficult for people to correctly perceive exponential growth \citep{wagenaar1975misperception, wagenaar1979pond} leading people to vastly underestimate exponential growth \citep{levy2017exponential}}. We provide a video with a linear y-scale online and show a snapshot of the video in Figure~\ref{fig-ch8:snapshot-linear-GW-numbers}.

\begin{figure*}
    \centering
\includegraphics[width=1\textwidth]{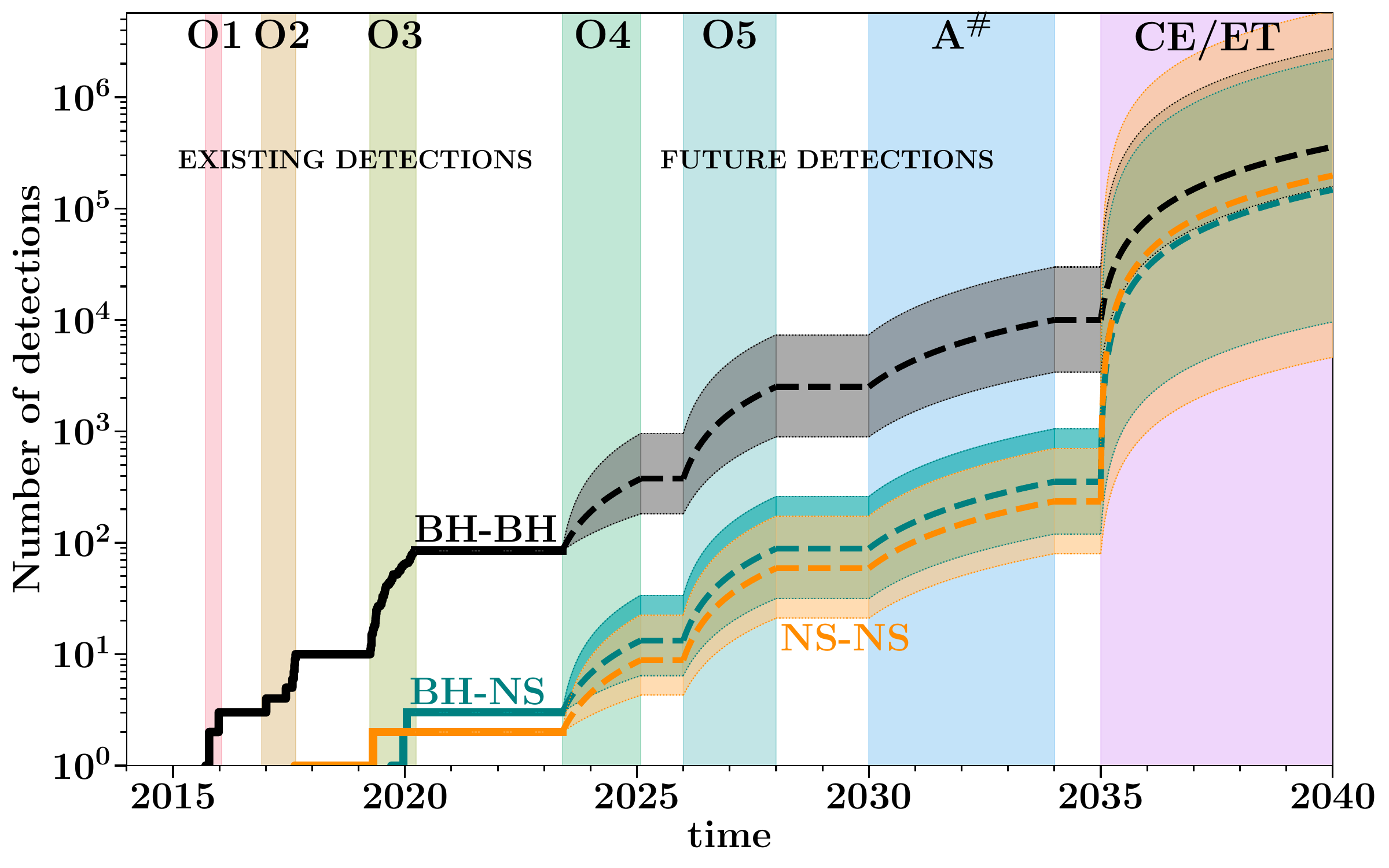} 
    \caption{Landscape of the number of detected mergers from binary black holes (BHBH; black), black hole-neutron stars (BHNS; teal), and binary neutron stars (NSNS; orange). See text for more details.   {The animation shows the catalog as a function of time and is available at \citet{zenodo:GWvisualizations}.}}  
    \label{fig-ch8:expected-detections-different-pipelines}
\end{figure*}

\begin{figure*}
    \centering
\includegraphics[width=1\textwidth]{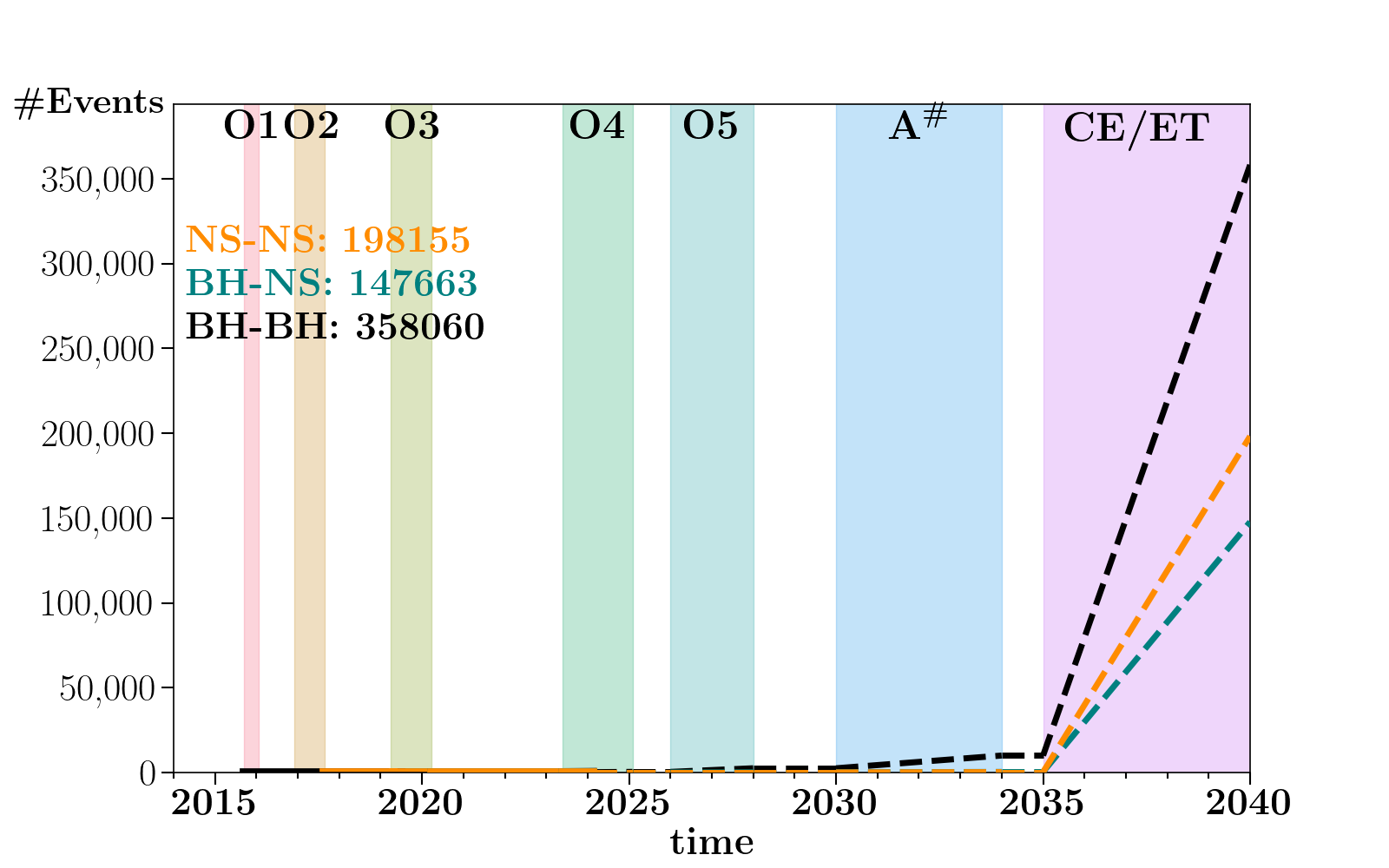} 
    \caption{Still of our animation that visualizes the growth of the gravitational-wave catalog in the coming decades on a linear scale.  Available at \citet{zenodo:GWvisualizations}. The linear y-scale emphasizes how the detections expected in the next decades will completely dwarf the number of detections observed today. }  
    \label{fig-ch8:snapshot-linear-GW-numbers}
\end{figure*}

\subsection{Expected Detections for O4 and O5}

The next upcoming LVK observing runs are O4 in 2023 and O5 around 2027. 
O4 started on May 24, 2023, and it is planned to have a total of 18-calendar month observing run, and LIGO, Virgo, and KAGRA detector sensitivities for binary neutron stars of about $160\,\rm{Mpc}$, $40$--$80\,\rm{Mpc}$, and $1$--$10\,\rm{Mpc}$ respectively\footnote{See \url{https://observing.docs.ligo.org/plan/index.html} for the most up-to-date information.}. 
O5 will start around 2027 and is planned to run for about two years with a targeted LIGO, Virgo, and KAGRA sensitivities of $240$--$325\,\rm{Mpc}$, $150$--$260\,\rm{Mpc}$, and $25$--$128\,\rm{Mpc}$, respectively. 

We discuss three recent publications reporting the number of detections expected for O4 and O5.
{These expected number of detections are based on an SNR threshold of 10, rather than the FAR or $\pastro{}$ for simplicity and ease of projecting onto future detector network sensitivities. This SNR threshold is expected to be equivalent to a FAR $\lesssim$ 1 per year (see the discussion about comparing O1, O2, and O3 injections in~\cite{Abbott:2021GWTC3}).}
First, the LVK Collaboration reports the prospects for the number of detections in O4 and O5 in \citet{Aasi:2013prospects}\footnote{Where we use the updated information for O4 and O5 from \href{https://emfollow.docs.ligo.org/userguide/capabilities.html}{\url{https://emfollow.docs.ligo.org/}}.}. Assuming among other things that the mass distributions follows the ``Power law $+$ Dip $+$ Break'' model from \citet{Abbott:2021GWTC3pop, Farah:2022}, they find an expected number of detections per year in O4 of about 11 \ac{NSNS}, 2 \ac{BHNS}, and 77 \ac{BHBH} mergers\footnote{Where we have scaled down the median numbers in the summary Table from \href{https://emfollow.docs.ligo.org/userguide/capabilities.html}{\url{https://emfollow.docs.ligo.org/}} by a factor 3.375 and rounded the value to go from public alert count to an approximate number of confident detections, see their footnote.}. For O5 the expected numbers are about 53 \ac{NSNS}, 9 \ac{BHNS}, and 258 \ac{BHBH} mergers. For both O4 and O5, the expected numbers have uncertainties that typically are of order unity or twice unity, reflecting both the merger rate uncertainty and the Poisson counting uncertainty (i.e. the probability of a given number of events occurring in a fixed interval of time). See also the discussion in \citet{2022ApJ...924...54P} for more detail. 
 Second, in the recent work by \citet{Iacovelli:2022} the authors present expectations for the number of detections per year in O4 in their Tables 3, 4, and 5 for \ac{BHBH}, \ac{BHNS}, and \ac{NSNS} detections. We add these also to Table~\ref{table:detectionsSummaryO4O5}. 
 Third, in \citet{Kiendrebeogo:2023} the authors perform a detailed analysis for O4 and O5 with based on two different assumptions for the underlying BHBH, BHNS and NSNS population (called LRR and GWTC-3). We add their rates for O4 and O5 from their Table~3. 

\begin{table*}
\caption{Summary of expected annual detection numbers for O4 and O5. Acronyms are: FAR (False Alarm Rate), GWTC (Gravitational Wave Transient Catalog), GWOSC (Gravitational Wave Open Science Center). For simplicity we do not show the provided uncertainties that often exceed unity.
References: \\
\footnotesize{(a) taken from \citet{Aasi:2013prospects} using their updated table from \href{https://emfollow.docs.ligo.org/userguide/capabilities.html}{\url{https://emfollow.docs.ligo.org/}}  and scaling down by a factor 3.375 to convert to confident detections  (representing SNR$\gtrsim 12$; corresponding to a single-detector SNR threshold of 8). \\ (b) Numbers are taken from Tables 3, 4, and 5 from \citet{Iacovelli:2022}. \\ (c) Numbers are taken from Table 3 from \citet{Kiendrebeogo:2023}.}
}
\label{table:detectionsSummaryO4O5}
	\centering
\begin{tabular}{@{}lccccc@{}}
	\toprule
	Label & Total &  BHBH & BHNS & NSNS & Ref \\ 		
	    %
	\hline \hline
	{\it \bf Detections for O4}  &  &  &  &  & \\
	\hline
        {\citet{Aasi:2013prospects} PBL model:}                    & \textbf{90}  & \textbf{77} & \textbf{2} & \textbf{11}      & {(a)} \\
        \hline 
        {\citet{Iacovelli:2022}:}                    & \textbf{90}  & \textbf{86} & \textbf{3} & \textbf{4}      & {(b)} \\
        \hline 
        {\citet{Kiendrebeogo:2023} LRR:}                    & \textbf{73}  & \textbf{46} & \textbf{10} & \textbf{17}      & {(c)} \\ \hline
        {\citet{Kiendrebeogo:2023} GWTC-3:}                    & \textbf{302}  & \textbf{260} & \textbf{6} & \textbf{36}      & {(c)} \\
        \hline 		\hline 	
	{\it \bf Detections for O5}  &  &  &  &  & \\
	\hline
        {\citet{Aasi:2013prospects} PBL model:}                    & \textbf{320}  & \textbf{258} & \textbf{9} & \textbf{53}      & {(a)} \\
        \hline 
        {\citet{Kiendrebeogo:2023} LRR:}                    & \textbf{324}  & \textbf{190} & \textbf{48} & \textbf{86}      & {(c)} \\ \hline
        {\citet{Kiendrebeogo:2023} GWTC-3:}                    & \textbf{1081}  & \textbf{870} & \textbf{31} & \textbf{180}      & {(c)} \\

        %
	%
	\hline 
	%
	%
	%
\bottomrule
\end{tabular}
\end{table*}

It is important to realize that the exact number of \ac{BHBH}, \ac{BHNS}, and \ac{NSNS} mergers that will be detected in O4 and O5 is highly uncertain. This number will depend on many factors including the actual sensitivity and observing duty cycle of the instruments and the Poisson statistics of the events. 
Another important uncertainty is the underlying population properties. 
One example is the  mass distributions of the double compact object merger populations. Since the detector sensitivity scales strongly with the source masses, the number of detected events is currently heavily biased towards more massive events, making the frequency and properties of lower-mass black holes more uncertain \citep{Abbott:2021GWTC3pop}. This makes it challenging to estimate the number of detections for future observing runs that probe larger distances and therefore see more of the underlying astrophysical population. Another important uncertainty is the redshift evolution of the rate and masses of the compact object merger events. From theory and observations, it is thought that the rate and properties of the double-compact objects change at higher redshifts \citep[e.g.][]{vanSon:2022, Abbott:2021GWTC3pop, Callister:2023arXiv230207289C}. Moreover,  different formation channels with different properties might be at play \citep{Ng:2021, Zevin:2021}. This can all impact the number of detections in future detectors. 

Given the many uncertain factors that can impact the number of detected events in future detectors, we decide to take a very simplistic approach for the purpose of the visualizations presented here. 
We caveat to the reader that these visualizations are to sketch a potential landscape of the number of expected gravitational-wave detections and that the actual expected number is still highly uncertain and requires detailed analysis. 
We instead simply use the LVK GWTC (GWOSC) catalog as a proxy for the ratio between \ac{BHBH}, \ac{BHNS}, and \ac{NSNS} detections as $85:3:2$, and assume a total number of 75 \ac{BHBH} mergers to be somewhat representative for a 11-month observing run at O3 sensitivity\footnote{Where 75 was chosen to exclude the 10 events from O1 and O2}. 
To obtain the expected number of detections for O4 and O5, we then scale the O3 detection number to the larger detection volume by using the LIGO O4 and O5 NSNS ranges and multiplying them by the planned observing run periods of 18 months and 2 years for O4 and O5, respectively. We also assume a uniform distribution of sources in an Euclidean universe. This gives an extremely rough estimate for O4 and O5 for BHBH sources of 
\begin{align*}
    N_{\rm{in \, O4}} &= 75 \times ({18}/{11}) \times (160/120)^3   = 291 \rm{\ total \ BHBH \ detections \ in \ 18 \ months,} \\
        N_{\rm{in \, O5}} &= 75 \times ({24}/{11})  \times (282.5/120)^3    = 2135 \rm{\ total \ BHBH \ detections  \ in \ 24 \ months}, 
\end{align*}
with $120\,\rm{Mpc}$ as proxy for the LIGO O3 mean \ac{NSNS} distance, $160\,\rm{Mpc}$ as a proxy for the average LIGO NSNS distance during O4, and $282.5\,\rm{Mpc}$ being the mean of the estimated LIGO \ac{NSNS} distance for O5\footnote{See \url{https://observing.docs.ligo.org/plan/}, accessed on 17/01/2024; the NSNS ranges could be updated over time.} Doing the same for the \ac{BHNS} and \ac{NSNS} rates using 3 \ac{BHNS} and 2 \ac{NSNS} detections in O3 gives 10 (75) BHNS and 7 (50) BNS expected in O4 (O5). 
For the uncertainty, we simply multiply and divide these numbers by a factor of three. 
We note that the numbers quoted in Table~\ref{table:detectionsSummaryO4O5} fall within our visualized uncertainty ranges, especially when including the uncertainties discussed in the studies (see \citealt{Kiendrebeogo:2023} for a detailed example of the effect of underlying assumptions).

\subsection{Expected Detections for $\rm{A}^{\sharp}$, Cosmic Explorer, and Einstein Telescope}
\begin{table*}
\caption{Summary of annual detection numbers expected with next-generation ground-based detectors Einstein Telescope and Cosmic Explorer (and variations). Acronyms are: FAR (false alarm rate), GWTC (Gravitational Wave Transient Catalog), GWOSC (Gravitational Wave Open Science Center, ET (Einstein Telescope), CE (Cosmic Explorer) N/A (not applicable; authors did not mention rates for this flavor). 
References: \\
\footnotesize{(a) Taken from the minimum and maximum numbers for ET-B and CE from Table 2 in \citet{Baibhav:2019}.\\
(b) Taken from Table~4 from \citet{Borhanian:2022arXiv220211048B} based events with  SNR $\gtrsim 10$. We take the lowest and highest (ECS) rates for the pessimistic and optimistic rates, relatively. \\(c) Taken from Tables~3, 4, and 5 from \citet{Iacovelli:2022}, where we use the `Analysed' events that have SNR$\gtrsim 12$. \\(d) Taken from the lowest (pessimistic) and highest (optimistic) bounds in Section 5.4 from \citet{Iacovelli:2022}. \\
(e) Taken from Table~6 in {\citet{Ronchini:2022}}. \\ (f) Taken from column 4 (events with  SNR$\gtrsim 10$) in Table~IV in {\citet{Gupta:2023}}. \\  We include error bars and take the lowest and highest values for the pessimistic and optimistic detection numbers, respectively. }
}
\label{table:detectionsSummaryCEandET}
	\centering
\begin{tabular}{@{}lccccc@{}}
	\toprule
	Label & Total &  BHBH & BHNS & NSNS & Ref \\ 		
	    %
	\hline \hline
	{\it \bf Detections for ET/CE}  &  &  &  &  & \\
	\hline
        {\citet{Baibhav:2019}  min (ET/CE)}                &   $\bf  5.3  \times 10^4 $     & $\bf 4.9  \times 10^4 $ & $\bf 2.4 \times 10^3 $ & $\bf 1.1 \times 10^3 $      & {(a)} \\
        {\citet{Baibhav:2019} max (ET/CE)}                &   $\bf   9.5 \times 10^5 $     & $\bf 5.4  \times 10^5 $ & $\bf 1.4 \times 10^5 $ & $\bf 2.7 \times 10^5 $      & {(a)} \\
        {\citet{Borhanian:2022arXiv220211048B} Pessimistic (HLKI$+$E):}                &    $\bf 1.4 \times 10^5 $     & $\bf 9.8 \times 10^4 $   & N/A &    $\bf 4.2 \times 10^4 $    & {(b)} \\
        {\citet{Borhanian:2022arXiv220211048B} Optimistic (ECS):}                &    $\bf 3.9 \times 10^5 $     & $\bf 1.2 \times 10^5 $   & N/A &    $\bf 2.7 \times 10^5 $    & {(b)} \\
        {\citet{Iacovelli:2022} (ET):}                          & $\bf 6.7 \times 10^4 $  & $\bf 4.8 \times 10^4 $  & $\bf 9.4 \times 10^3 $  & $\bf 9.2 \times 10^3 $      & {(c)} \\
        {\citet{Iacovelli:2022} (ET$+$2CE):}                    & $\bf 1.4 \times 10^5 $    & $\bf 7.0 \times 10^4 $  & $\bf 2.9 \times 10^4 $  & $\bf 4.0 \times 10^4 $      & {(c)} \\
        {\citet{Iacovelli:2022} Pessimistic:}                  &   $\bf 3.4 \times 10^4 $ & $\bf 3.1 \times 10^4 $ & $\bf 1.9 \times 10^3 $  & $\bf 9.1 \times 10^2 $     & {(d)} \\
        {\citet{Iacovelli:2022} Optimistic:}                &   $\bf 7.9 \times 10^5 $     & $\bf 1.1 \times 10^5 $ & $\bf 3.4 \times 10^4 $ & $\bf 6.5 \times 10^5 $      & {(d)} \\
        {{\citet{Ronchini:2022}} (ET only):}                &  N/A    &  N/A  & N/A  & $\bf 1.4 \times 10^5 $   & {(e)} \\
        {{\citet{Ronchini:2022}} (ET $+$ CE):}                &  N/A    &  N/A  & N/A  & $\bf 4.6 \times 10^5 $   & {(e)} \\
        {{\citet{Ronchini:2022}} (ET $+$ 2CE):}                &  N/A    &  N/A  & N/A  & $\bf 5.9 \times 10^5 $    & {(e)} \\
        {\citet{Gupta:2023} Pessimistic:}                  &   $\bf 7.6 \times 10^4 $ & $\bf 5.0 \times 10^4 $ & $\bf 9.0 \times 10^3 $  & $\bf 1.7 \times 10^4 $     & {(f)} \\
        {\citet{Gupta:2023} Optimistic:}                &   $\bf 1.8  \times 10^6 $     & $\bf 1.5 \times 10^5 $ & $\bf 4.4 \times 10^5 $ & $\bf 1.2 \times 10^6 $      & {(f)} \\
        \hline 
        
        minimum &   $ 3.4 \times 10^4 $  &   $ 3.1 \times 10^4 $  &   $ 1.9 \times 10^3 $  &   $ 9.1 \times 10^2 $ &  \\
        maximum &   $ 1.8 \times 10^6 $  &   $ 5.4 \times 10^5 $  &   $ 4.4 \times 10^5 $  &   $ 1.2 \times 10^6 $ &  \\
        \hline 
        %
	%
	%
	%
\bottomrule
\end{tabular}
\end{table*}

The exact sensitivity, and hence the expected number of detections that will be made with future upgrades to the LIGO and Virgo detectors  beyond O5 is uncertain\footnote{See \href{https://dcc.ligo.org/public/0183/T2200287/002/T2200287v2_PO5report.pdf}{T2200287v2\_PO5report.pdf} for an overview of proposed detector network scenarios.}. In \citet{Baibhav:2019, Voyager:2020,  Gupta:2023}, the authors mention that a future best sensitivity of the LIGO/Virgo detectors, sometimes called $\rm{A}^{\sharp}$ (or Voyager) might detect about a hundred times more sources compared to Advanced LIGO/Advanced Virgo/KAGRA (sometimes referred to as $\rm{A}^{+}$/$\rm{AdV}^{+}$). For simplicity, we decide to use a factor of about 100 times more sources in $\rm{A}^{\sharp}$ compared to our O3 estimate in Figure~\ref{fig-ch8:expected-detections-different-pipelines}. We also include a factor 3 for the uncertainty estimate.  

The number of detections expected with Cosmic Explorer (CE) and Einstein Telescope (ET) is also still highly uncertain. A source of uncertainty includes the (local) merger rate, which is used to calculate the underlying merger rate population. We include for the visualizations the following recent work calculating the expected number of detections. 
\citet{Gupta:2023} present expected detection numbers for GW networks from $\rm{A}^{\sharp}$ to networks with three next-generation detectors. We take their highest and lowest detection numbers for networks with at least one next-generation detector from their Table IV and add it to table~\ref{table:detectionsSummaryCEandET} as a proxy for a CE/ET era detector. 
 In \citet{Iacovelli:2022} present expectations for the number of  \ac{BHBH}, \ac{BHNS}, and \ac{NSNS} detections with an ET or ET and two CE-like interferometer networks (see their Tables 3, 4, and 5). We add these rates to Table~\ref{table:detectionsSummaryCEandET}. 
In Section~5.4 \citet{Iacovelli:2022} discuss the impact on the detection number by changing the rate of the underlying population. We take the most optimistic and pessimistic numbers from these ranges and add this also to Table~\ref{table:detectionsSummaryCEandET}. Expected numbers with next-generation detectors have also been calculated by \citet{Borhanian:2022arXiv220211048B, Ronchini:2022}, whose thorough analysis of the prospects for CE and ET are aligned with the bounds by \citet{Iacovelli:2022} and \citet{Gupta:2023}, and we add their rates to Table~\ref{table:detectionsSummaryCEandET}. \citet{Baibhav:2019} calculated expected detection rates for BHBH, BHNS, and NSNS for future detector networks based on population synthesis models. We add their minimum and maximum expected rates for an ET or CE detector to Table~\ref{table:detectionsSummaryCEandET}. 
The authors in these papers use slightly different models for the local merger rates, detection threshold (SNR), detector network, and underlying population, leading to the different answers, see appendix B in \citet{Iacovelli:2022} for a detailed discussion. 
For our visualizations in Figure~\ref{fig-ch8:expected-detections-different-pipelines}
we use the minimum and maximum detection numbers from the values above, as shown in the last two rows in Table~\ref{table:detectionsSummaryCEandET}. These numbers are aligned with expectations from earlier work such as \citet{Baibhav:2019}.

\section{Discussion and Conclusion}

\subsection{Caveats}
{
For visualization purposes, we focused in this paper on simplified approximations for the (expected) detected number of \ac{GW} sources. For example, the studies used for the visualization  (Tables~{\ref{table:detectionsSummary}, \ref{table:detectionsSummaryO4O5}, \ref{table:detectionsSummaryCEandET}}) often use different SNR or FAR thresholds and different underlying population properties --- sometimes even within the same catalog. As a result, we note that the visualization in Figure~\ref{fig-ch8:expected-detections-different-pipelines} uses different thresholds for the existing and future detections. This has minimal impact on the visualization given the many other uncertainties. 
We provide an extra version of Figure~\ref{fig-ch8:expected-detections-different-pipelines} in our online repository  that for the detections in O1, O2, and O3 uses GWTC-3 (GWOSC)  events with `matched-filter' SNR $\geq 10$ (which contains 55 BBH, 1 BHNS, and 2 BNS mergers) instead of the $\pastro>0.5$ threshold in Figure~\ref{fig-ch8:expected-detections-different-pipelines} to align more with the threshold typically used in the studies for future detections (see also Section~\ref{sec:intro-to-detection-number-and-catalog-construction}).
Moreover, the detection analyses don't account for uncertainties in the population distribution of mergers. Additionally, with multiple pipelines analyzing the same underlying \ac{GW} data, there are non-trivial trial factors that most analyses do not fully account for. There is significant ongoing work in the \ac{GW} community on making these statistical foundations more robust; by better measuring our sensitivity to signals~\citep{Essick:2023, EssickFischbach:2023, Talbot:2023pex}, by combining and comparing different catalogs in a more self-consistent way~\citep{Sutton:2009, Biswas:2012ty, Banagiri:2023, Mould:2023}, and by development of techniques for simultaneous inference of detection statistics and population properties~\citep{Roulet:2020wyq, Galaudage:2019jdx}. In addition, future work should investigate in more detail the effect on the number of detections from uncertain source classification (i.e. whether a detection classified as a BHBH could be a BHNS or NSNS). A larger number of detected events will help increase our understanding of the underlying merger population and guide individual source classification \citep{Abbott:2021GWTC3pop}.}

\subsection{Future Outlook: beyond detector numbers}
Future detectors such as $\rm{A}^{\#}$, Einstein Telescope, and Cosmic Explorer will transform astronomy much beyond just increasing the number of detections, which has been the focus of this paper. 
{While our understanding of the population improves as the square-root of the number of observations,} these next-generation detectors provide data with ever-increasing precision, at wider accessible gravitational-wave frequency bands, and out to larger distances, allowing detections of gravitational-wave sources possibly beyond redshifts of ten. This will allow us to address the redshift evolution of these sources, and probe compact objects and their progenitor (stars) in environments vastly different from the Milky Way.  
Moreover, an important development is that future detectors beyond increasing detection numbers will also detect many more events with higher precision (SNR), which will allow scientists to infer the source properties with more precision (see e.g., the discussions in \citet{2012PhRvD..86l2001R, Abbott:2021-GWTC-2-1, Borhanian:2022arXiv220211048B, Iacovelli:2022, Ronchini:2022,  Evans:2023}), which is important to address many open questions. For example. future detectors could determine the neutron star equation of state \citep[e.g.,][]{Coupechoux:2023arXiv230204147C}, can provide early warning detections of binary neutron star mergers \citep[e.g.,][]{Magee:2022ApJ...935..139M}, provide improved sky localization allowing for more multi-messenger observations \citep[e.g.,][]{2018PhRvD..97j4064M, 2022ApJ...924...54P}, might be able to observe the existence of black holes before star formation took place \citep[e.g.,][]{Chen:2020JCAP...08..039C, Ng:2022ApJ...931L..12N}, and can provide improved tests of General Relativity and estimates for the graviton mass \citep[e.g.][]{Perkins:2021}. We refer the reader to \citet{Reitze:2019, Maggiore:2020EinsteinTelescope, Evans:2021arXiv210909882E, Evans:2023}, and \citet{Gupta:2023} and the references therein for more details and a detailed description of the science case for next-generation detectors. 

All in all, we present in this paper publicly available visualizations to highlight the many detections to date in gravitational-wave astronomy as well as the expected detections in the exciting decades to come!

\section{Software and third party data repository citations} \label{sec:cite}
All data and code to reproduce the results, visualizations, and figures in this paper are publicly available at \citet{zenodo:GWvisualizations}. This code makes use of the publicly available GWOSC catalog from \citet{GWOSC:2023paper, GWOSC:dataset}, the 4-OGC catalog from \citet{Nitz:2023-4-OGC}, and the GWTC-3 population catalogs based on \citet{Abbott:2021GWTC3pop} retrieved from \citet{Zenodo:GWTC-3pop}. The IAS data is based on \citet{Venumadhav:2019, Venumadhav:2020, 2019PhRvD.100b3007Z,  Olsen:2022, mehta2023new}, and  \citet{Wadekar:2023}; see references in the papers for the repositories of the underlying IAS data. The code made use of \textsc{Python} from the Python Software Foundation available at \url{http://www.python.org} \citep{CS-R9526}. In addition, the following Python packages were used: \textsc{Matplotlib} \citep{2007CSE.....9...90H},  \textsc{NumPy} \citep{2020NumPy-Array}, \textsc{SciPy} \citep{2020SciPy-NMeth}, \textsc{IPython$/$Jupyter} \citep{2007CSE.....9c..21P, kluyver2016jupyter}, 
\textsc{Astropy} \citep{2018AJ....156..123A}  and   \href{https://docs.h5py.org/en/stable/}{\textsc{hdf5}} \citep{collette_python_hdf5_2014}. 
}

\begin{acknowledgments}
 FSB is grateful for the many suggestions and feedback received for this work from a list of people including Will Farr,  Edo Berger, Carl-Johan Haster, Karan Jani, Alex Nitz, Soheb Mandhai, Eric Thrane, Ilya Mandel, Yvette Cendes, Maryam Hussaini, and Lieke van Son. All authors especially like to thank Christopher Berry for many discussions and feedback.
{We thank the anonymous referee for the insightful and valuable comments on the manuscript.}
FSB was supported by NASA FINESST scholarship 80NSSC22K1601, awarded to the President and Fellows of Harvard College. FSB is also grateful for the support on visualizing research received through the Harvard Horizons Scholarship program. FSB is also supported by Simons Foundation award 1141468 hosted at Columbia University. 
{LIGO was constructed by the California Institute of Technology and Massachusetts Institute of Technology with funding from the National Science Foundation and operates under cooperative agreement PHY-1764464 . This paper carries LIGO Document Number LIGO-P2300435}.
This research has made use of data or software obtained from the Gravitational Wave Open Science Center (gwosc.org), a service of the LIGO Scientific Collaboration, the Virgo Collaboration, and KAGRA. This material is based upon work supported by NSF's LIGO Laboratory which is a major facility fully funded by the National Science Foundation, as well as the Science and Technology Facilities Council (STFC) of the United Kingdom, the Max-Planck-Society (MPS), and the State of Niedersachsen/Germany for support of the construction of Advanced LIGO and construction and operation of the GEO600 detector. Additional support for Advanced LIGO was provided by the Australian Research Council. Virgo is funded, through the European Gravitational Observatory (EGO), by the French Centre National de Recherche Scientifique (CNRS), the Italian Istituto Nazionale di Fisica Nucleare (INFN) and the Dutch Nikhef, with contributions by institutions from Belgium, Germany, Greece, Hungary, Ireland, Japan, Monaco, Poland, Portugal, Spain. KAGRA is supported by Ministry of Education, Culture, Sports, Science and Technology (MEXT), Japan Society for the Promotion of Science (JSPS) in Japan; National Research Foundation (NRF) and Ministry of Science and ICT (MSIT) in Korea; Academia Sinica (AS) and National Science and Technology Council (NSTC) in Taiwan.
\end{acknowledgments}

%





\bibliography{biblio,biblio2,biblio3}{}
\bibliographystyle{aasjournal}



\end{document}